\documentclass{PoS}
\title{A Monte Carlo analysis of the SMEFT in the top quark sector}

\ShortTitle{A Monte Carlo analysis of the SMEFT in the top quark sector}

\author{\speaker{Emma Slade} \\
Rudolf Peierls Centre for Theoretical Physics, University of Oxford, \\
 Clarendon Laboratory, Parks Road, Oxford OX1 3PU, United Kingdom \\
 
E-mail: \email{emma.slade@physics.ox.ac.uk}}

\abstract{We present a framework for carrying out global analyses of the
Standard Model Effective Field Theory: {SMEFiT}.
This approach is based on the Monte Carlo replica method, widely used in the case of NNPDF fits of the proton structure, for deriving a
faithful estimate of the experimental and theoretical uncertainties.
As a proof of concept of the SMEFiT methodology,
we present a study of the constraints on the SMEFT
provided by top quark production measurements from the LHC.
We derive bounds for the 34 degrees of freedom relevant for
the interpretation of the LHC top quark data and compare these bounds
with previously reported constraints.
}

\FullConference{XXVII International Workshop on Deep-Inelastic Scattering and Related Subjects - DIS2019\\
		8-12 April, 2019\\
		Torino, Italy}

\begin{document}
\paragraph{Introduction}
The Large Hadron Collider (LHC) is pursuing an extensive
program of direct searches for physics beyond the
Standard Model (BSM) by exploiting its unique reach in energy.
As well as searching for the direct production
of new particles, one can also perform indirect searches,
where precise measurements are compared to Standard Model (SM) predictions with the aim of discovering BSM effects via deviations in the tails of SM distributions.

A powerful framework with which we can parameterise these deviations with respect to the SM in a model-independent way is
the Standard Model Effective Field Theory
(SMEFT)~\cite{Weinberg:1979sa,Buchmuller:1985jz,Grzadkowski:2010es}.
In the SMEFT, the effects of BSM dynamics at high scales $E\simeq \Lambda$ 
are parametrised for $E\ll \Lambda$ in terms of higher-dimensional (irrelevant)
operators built up from the SM fields and symmetries.
Analysing experimental data in the SMEFT framework is non-trivial as even when only considering operators that conserve
baryon and lepton number~\cite{Grzadkowski:2010es}, one ends
up with over 2000 operators at dimension-6 in the absence of flavour assumptions.
This implies that global SMEFT analyses need to explore 
a complicated parameter space with a large number of degenerate (``flat'') 
directions.

From the methodological point of view, a global fit of the SMEFT
from LHC measurements requires combining state-of-the-art theoretical
calculations (in the SM and in the SMEFT) with a wide variety of experimental 
cross-sections and distributions.
In this work we develop a novel
strategy for global SMEFT analyses.
This approach~\cite{Hartland:2019bjb}, which we denote by SMEFiT, combines the generation of 
Monte Carlo (MC) replicas to estimate and propagate uncertainties, with
 methodological techniques such as closure testing and cross-validation.

As a proof of concept of the SMEFiT methodology, we apply it here for the first 
time to the study of top quark production at the LHC in the SMEFT 
framework at dimension-6.
We include the NLO QCD corrections to the SMEFT
contributions and compute both the linear  ($\mathcal{O}(\Lambda^{-2})$) and the 
quadratic  ($\mathcal{O}(\Lambda^{-4})$) contributions to the SMEFT predictions.
By exploiting the SMEFiT methodology, we derive the probability
distribution in the space of SMEFT Wilson coefficients.

\paragraph{The SMEFT framework}

Let us begin by reviewing the SMEFT
formalism~\cite{Buchmuller:1985jz,Weinberg:1978kz},
with emphasis on its description of the top quark sector.
As mentioned above, the effects of new BSM particles with
a mass scale $M\simeq \Lambda$ can be
parametrised at lower energies $E\ll \Lambda$ in a model-independent way in
terms of a basis of operators constructed from the SM fields
and their symmetries.
The resulting Lagrangian then admits the following expansion
\begin{equation}
\label{eq:smeftlagrangian}
\mathcal{L}_{\rm SMEFT}=\mathcal{L}_{\rm SM} + \sum_i^{N_{d6}} \frac{c_i}{\Lambda^2}\mathcal{O}_i^{(6)} +
\sum_j^{N_{d8}} \frac{b_j}{\Lambda^4}\mathcal{O}_j^{(8)} + \ldots \, ,
\end{equation}
where $\mathcal{L}_{\rm SM}$ is the SM Lagrangian, and
$\{\mathcal{O}_i^{(6)}\}$ and $\{\mathcal{O}_j^{(8)}\}$ stand for
the elements of the
operator basis of mass-dimension $d=6$ and $d=8$,
respectively.
Operators with $d=5$ and $d=7$, which violate lepton and/or baryon number
conservation~\cite{Degrande:2012wf,Kobach:2016ami}, are not considered here.
In this work we adopt the Warsaw basis for
$\{\mathcal{O}_i^{(6)}\}$~\cite{Grzadkowski:2010es}, and neglect effects
arising from operators with mass dimension $d\ge 8$.

In general, the effects of the dimension-6 operators can be written as follows:
\begin{equation}
\label{eq:smeftXsecInt}
\sigma=\sigma_{\rm SM} + \sum_i^{N_{d6}}
\kappa_i \frac{c_i}{\Lambda^2} +
\sum_{i,j}^{N_{d6}}  \widetilde{\kappa}_{ij} \frac{c_ic_j}{\Lambda^4}  \, ,
\end{equation}
where $\sigma_{\rm SM}$ indicates
the SM prediction and $c_i$ are the Wilson coefficients we wish to fit.

In Eq.~(\ref{eq:smeftXsecInt}), the second term arises from operators 
interfering with the SM amplitude.
The resulting $\mathcal{O}(\Lambda^{-2})$ corrections to the SM 
cross-sections represent formally the dominant correction, though in many cases 
they can be subleading.
The third term in Eq.~(\ref{eq:smeftXsecInt}),
representing $\mathcal{O}(\Lambda^{-4})$ effects, are from the
squared amplitudes of the SMEFT operators.
In principle, this term may not need to be included, depending on whether the truncation at
$\mathcal{O}(\Lambda^{-2})$ order is done at the Lagrangian or the cross-section level, but in practice there are often
valid reasons to include them in the calculation.

In this work we follow the strategy
documented in the LHC Top Quark Working Group 
note~\cite{AguilarSaavedra:2018nen}.
We adopt the Minimal
Flavour Violation (MFV) hypothesis~\cite{DAmbrosio:2002vsn} in the quark
sector as the baseline scenario.
We further assume that the CKM matrix is diagonal, such that the Yukawa
couplings are non-zero only for the top and bottom quarks.
In other words, we
impose a $U(2)_q\times U(2)_u \times U(2)_d$ flavour symmetry in the first
two generations.
In addition, we restrict ourselves to the CP-even operators, and focus on
those operators that induce modifications in the interactions of the top quark
with other SM fields.
Under these assumptions, we end up with 34 independent degrees of freedom to fit.

\paragraph{Methodology}

In this work, we adopt the MC replica method, inspired by the NNPDF fits to parton densities (see Ref.~\cite{Ball:2017nwa} and references therein), to propagate the 
experimental uncertainties from  experimental cross-sections
to the fitted SMEFT coefficients $\{c_i\}$.
The idea is to construct a sampling of the probability
distribution in the space of the experimental data, which then translates 
into a sampling of the probability distribution in the space of the SMEFT 
coefficients.
This strategy can be implemented by generating a large number
of artificial replicas of the original data.
The main advantage of the MC method is that it does not make any 
assumption about the probability distribution of the coefficients, 
and is not limited to Gaussian distributions.
Moreover, it is suited to problems where the parameter space is large
and complicated, with a large number of quasi-degenerate minima and flat
directions.

In this work, we use as input to all our theory calculations the NNPDF3.1 NNLO
no-top PDF set~\cite{Ball:2017nwa}. By removing the top data from the PDF fit, this prevents us double-counting the data both in the PDFs and the SMEFT fits.
To account for the removal of the data in the PDF fit, we include PDF uncertainties in the  covariance matrix. As we do not currently
account for missing higher-order uncertainties, we use NNLO QCD predictions for all available SM processes, and NLO otherwise.

\paragraph{Results}
In the following, we present the fit results
for the central values $\left\langle c_ i \right\rangle$, defined as
\begin{equation}
\label{eq:meancoefficient}
\left\langle c_l\right\rangle \equiv \frac{1}{N_{\rm rep}}\sum_{k=1}^{N_{\rm rep}} c_l^{(k)}\, ,
\end{equation}
 and the corresponding 95\% CL uncertainties, $\delta c_i$,
for the 34 dimension-6 SMEFT degrees of freedom. In all cases, the number of replicas $N_{\mathrm {rep}} = 1000$.
We also study the cross-correlations between these degrees of freedom
as they provide an important piece of information
since  these correlations
might be large because of flat directions in the parameter space.

\begin{figure}[t]
  \begin{center}
\includegraphics[width=0.49\linewidth]{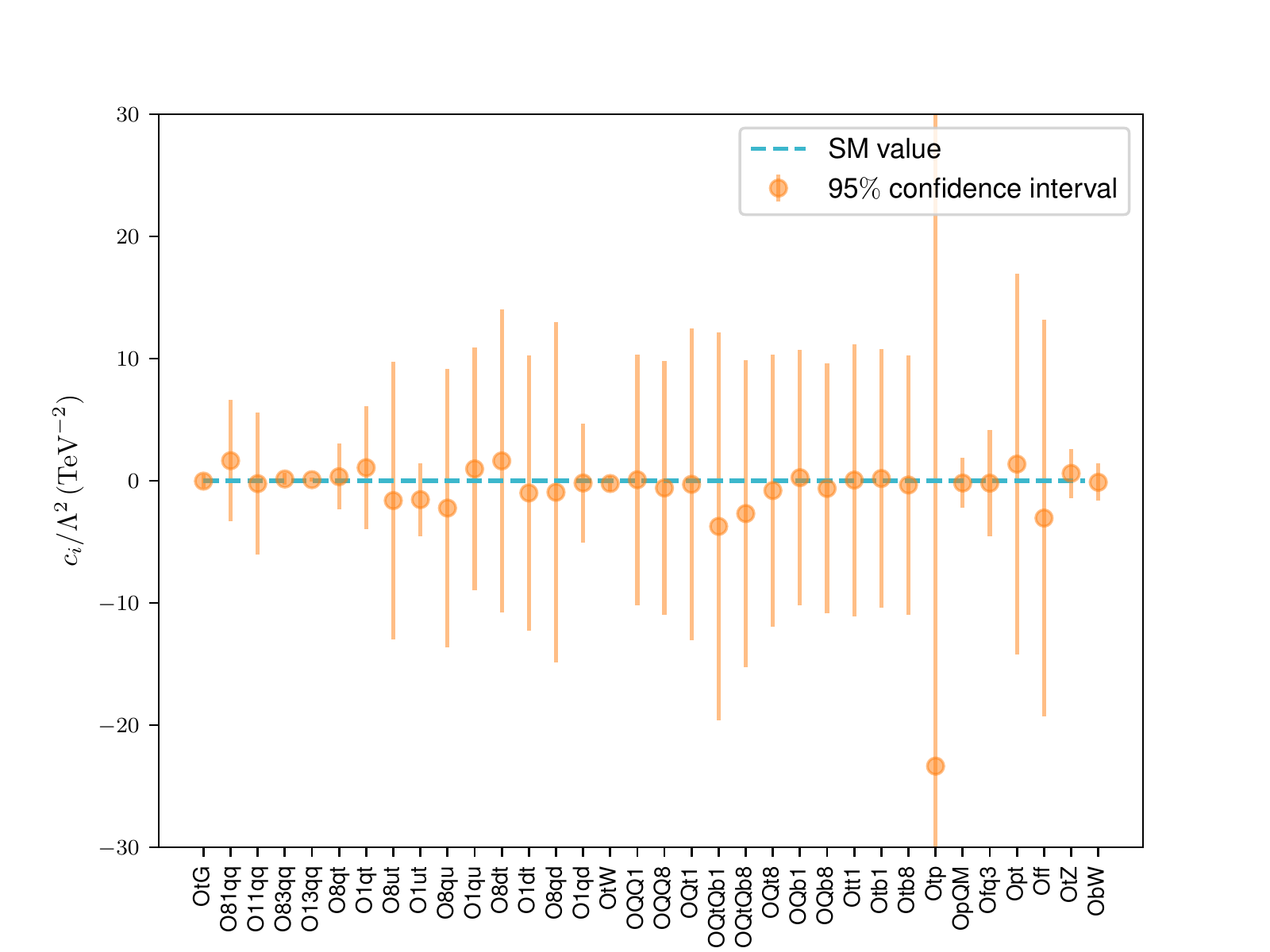}
\includegraphics[width=0.49\linewidth]{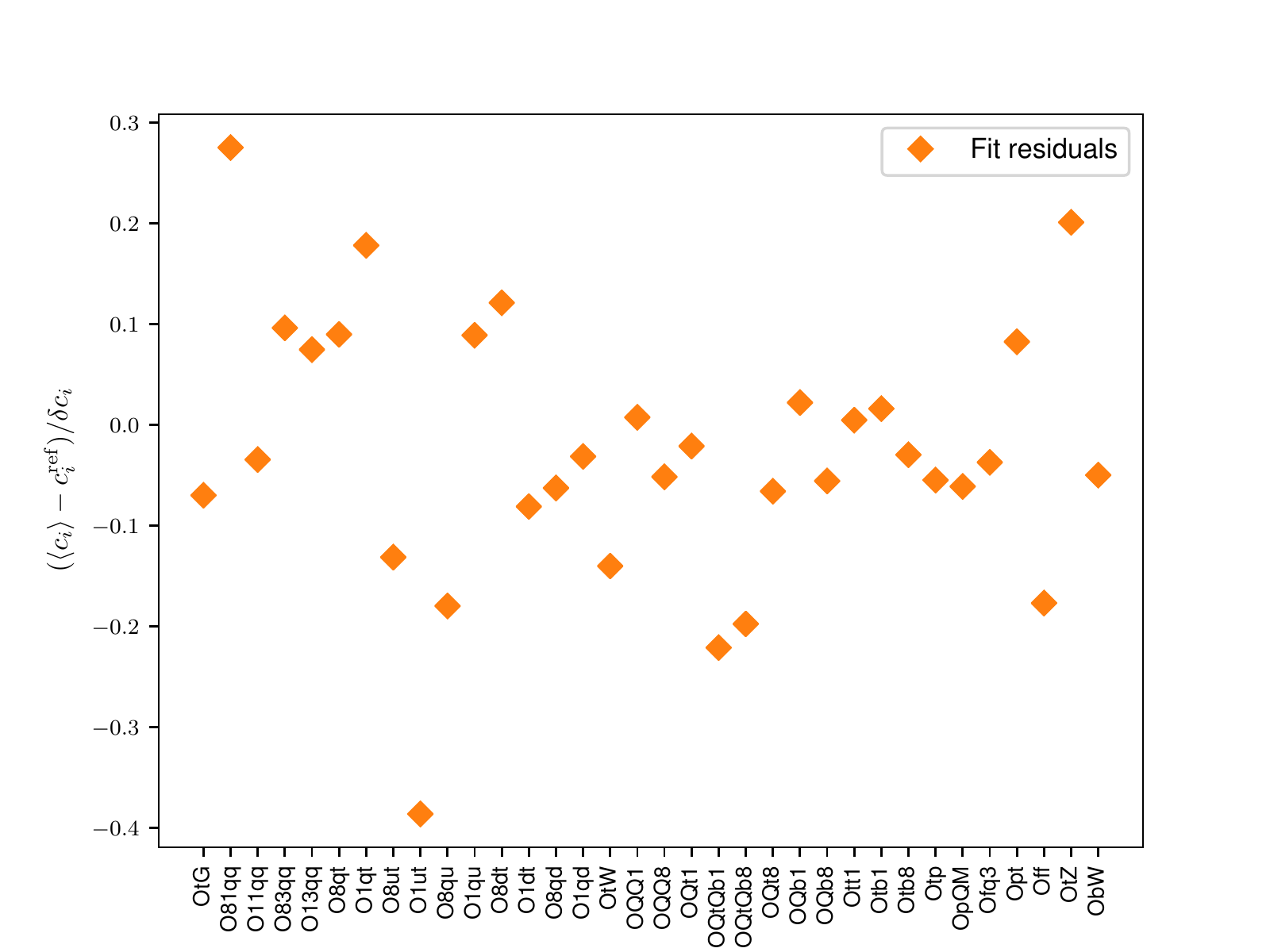}
\caption{\small Left: the best-fit values
  with the corresponding 95\% confidence intervals
  for the degrees of freedom considered in this
  analysis.
  Right plot: the associated fit residuals.
     \label{fig:SMEFT_std_dev_crossvalON_NLO_HO} }
  \end{center}
\end{figure}

In Fig.~\ref{fig:SMEFT_std_dev_crossvalON_NLO_HO} we display
the best-fit values with the corresponding
95\% confidence levels.
The dashed line indicates the SM prediction as reference.
In the right panel, we show the associated fit residuals $r_i$, 
\begin{equation}
\label{eq:fitresiduals}
r_i \equiv \frac{( \left\langle c_i\right\rangle - c_i^{\rm (ref)})}{\delta c_i} \, ,
\end{equation}
which measure the deviation
of the fit results with respect to the SM in units
of the 95\% CL uncertainties.

We find that
the fit results are in good agreement with the SM within
uncertainties.
Note that, due to the correlations between the degrees of freedom, the size of the residuals are smaller than in the case where they are completely independent. We have explicitly verified this; see~\cite{Hartland:2019bjb} for more details.
We also observe that
there are a wide range of values for the fit uncertainties for the different degrees of freedom.
The origin of these differences is due to the fact that different
degrees of freedom are constrained by different processes,
and in each case the amount of experimental information
varies quite considerably.

\begin{figure}[t]
  \begin{center}
\includegraphics[scale=0.6]{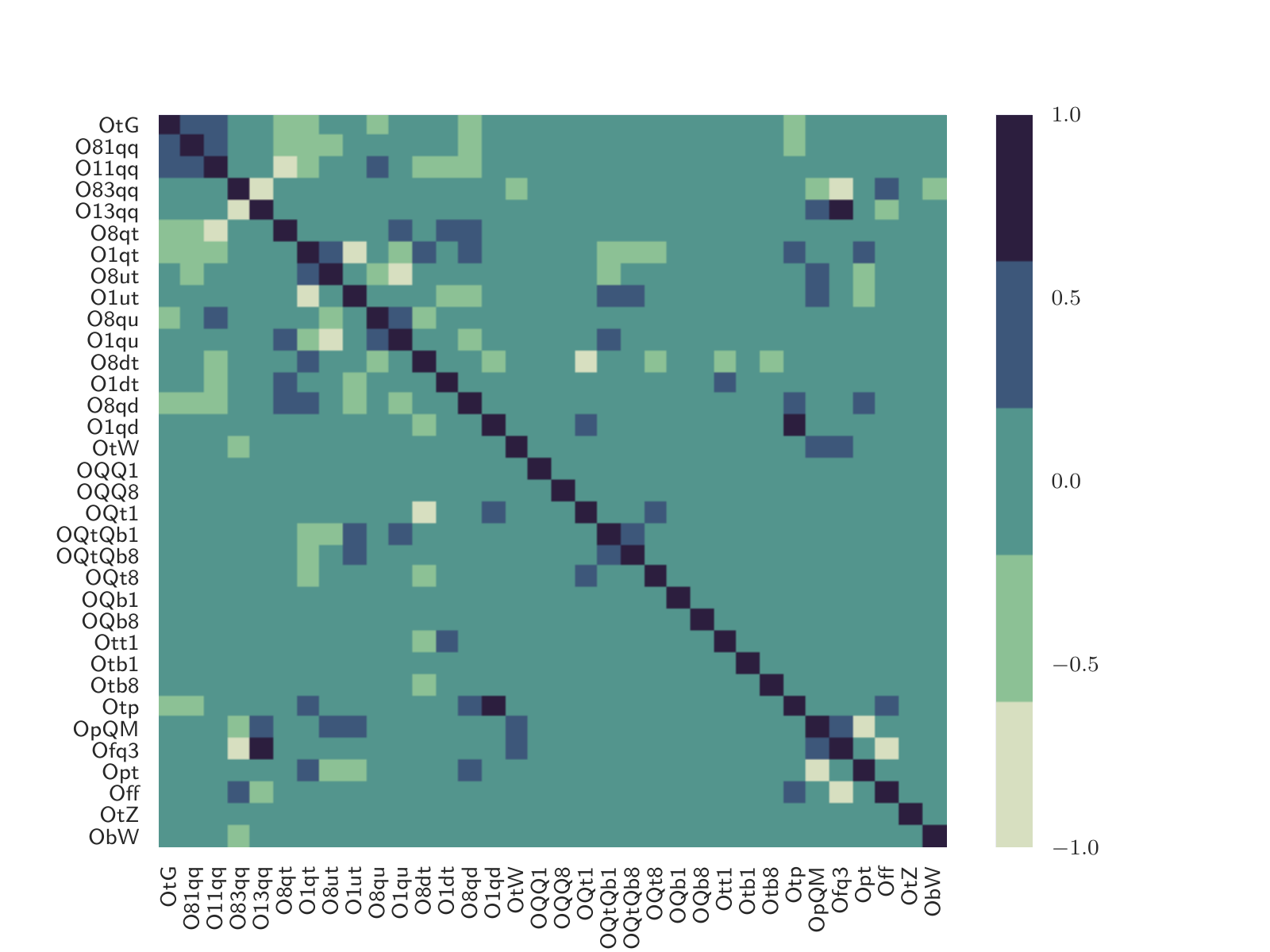}
\caption{\small Heat map indicating the values of the correlation coefficient
  $\rho(c_i,c_j)$
  between the 34 fitted coefficients
  shown in Fig.~\ref{fig:SMEFT_std_dev_crossvalON_NLO_HO},
  see text for more details.
     \label{fig:heatmap_crossvalON_NLO_HO} }
  \end{center}
\end{figure}

The interpretation of the 95\% CL uncertainties shown in 
Fig.~\ref{fig:SMEFT_std_dev_crossvalON_NLO_HO}, requires some care.
The reason is that, with the available experimental data, we are not able to
fully seperate the independent directions in the parameter space.
As a consequence,
there will be in general correlations between the fit
parameters.
To quantify this, in Fig.~\ref{fig:heatmap_crossvalON_NLO_HO} we show
a heat map indicating the values of the correlation coefficient,
\begin{equation}
\label{eq:correlationL2CT}
\rho( c_i,c_j)=\frac{\frac{1}{N_{\rm rep}}\sum_{k=1}^{N_{\rm rep}}
c_i^{(k)} c_j^{(k)} -\left\langle c_i\right\rangle \left\langle c_j\right\rangle
}{\delta c_i \delta c_j} \, .
\end{equation}
between the degrees of freedom constrained from the fit.
In this heat map, dark blue regions correspond to degrees of freedom
that are significantly correlated, while light green regions
are  degrees of freedom that are significantly anti-correlated.
The effects of such correlations are ignored in fits
where these degrees of freedom are constrained individually rather
than marginalised from the global fit results, and lead
in general to artificially tighter constraints.

\begin{figure}[t]
  \begin{center}
\includegraphics[width=0.99\linewidth]{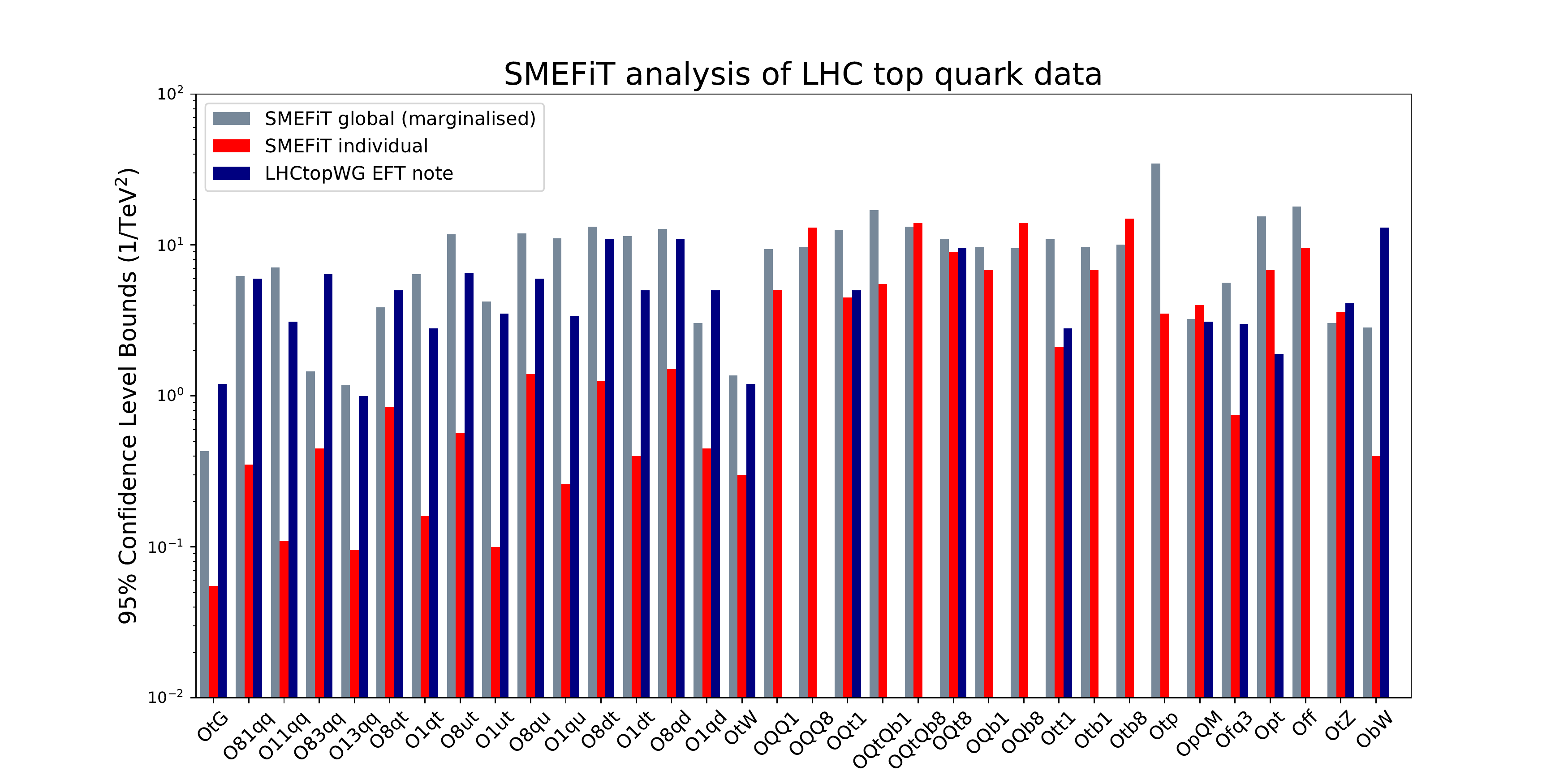}
    \caption{\small The 95\% CL bounds on the 34 degrees of freedom
      included the present analysis, both in the marginalised (global) 
      and in the individual fit cases, with the
      bounds reported in the LHC Top WG EFT note~\cite{AguilarSaavedra:2018nen}.
      \label{fig:SMEFiT-bounds}
  }
  \end{center}
\end{figure}

We finally show the comparison between
our global fit results and the bounds reported in
the LHC top WG EFT note, as well as with
 individual fit results in Fig.~\ref{fig:SMEFiT-bounds}.
We find that for some of the fitted degrees of freedom our bounds are stronger 
than those reported in previous studies, in some cases by 
nearly one order of magnitude.
One can see how the individual bounds
are in general tighter or at most comparable to the marginalised ones. 
This emphasises the point that individual fits to operators produce artifically
tight bounds. It is only in global fits, where the correlations are accounted for,
that we are able to produce realistic estimates.

\paragraph{Conclusion}
The study presented in this work is the first proof-of-principle application of the SMEFiT framework. 
As a proof-of-concept of the SMEFiT framework, we have presented an  analysis of top quark production measurements at the LHC at 8 TeV and 13 TeV.
Our results are in good agreement with the SM expectations: we find that all the fitted SMEFT degrees of freedom are consistent with the SM result at the 95\% CL.
In addition, SMEFiT is easily able to cope with enlarging the fitted parameter space by adding new experimental data.

It further allows for the easy addition of new data via Bayesian reweighting~\cite{Ball:2010gb,Ball:2011gg} without having to rerun fits or have direct access to the SMEFiT code. An open-source reweighting code which allows one to study the impact of new data on the baseline fit presented here will be released soon.

\providecommand{\href}[2]{#2}\begingroup\raggedright\endgroup


\begin{thebibliography}{10}

\bibitem{Weinberg:1979sa}
S.~Weinberg, {\it {Baryon and Lepton Nonconserving Processes}},  {\em Phys.
  Rev. Lett.} {\bf 43} (1979) 1566--1570.

\bibitem{Buchmuller:1985jz}
W.~Buchmuller and D.~Wyler, {\it {Effective Lagrangian Analysis of New
  Interactions and Flavor Conservation}},  {\em Nucl. Phys.} {\bf B268} (1986)
  621--653.

\bibitem{Grzadkowski:2010es}
B.~Grzadkowski, M.~Iskrzynski, M.~Misiak, and J.~Rosiek, {\it {Dimension-Six
  Terms in the Standard Model Lagrangian}},  {\em JHEP} {\bf 10} (2010) 085,
  [\href{http://arxiv.org/abs/1008.4884}{{\tt arXiv:1008.4884}}].

\bibitem{Hartland:2019bjb}
N.~P. Hartland, F.~Maltoni, E.~R. Nocera, J.~Rojo, E.~Slade, E.~Vryonidou, and
  C.~Zhang, {\it {A Monte Carlo global analysis of the Standard Model Effective
  Field Theory: the top quark sector}},  {\em JHEP} {\bf 04} (2019) 100,
  [\href{http://arxiv.org/abs/1901.05965}{{\tt arXiv:1901.05965}}].

\bibitem{Weinberg:1978kz}
S.~Weinberg, {\it {Phenomenological Lagrangians}},  {\em Physica} {\bf A96}
  (1979) 327.

\bibitem{Degrande:2012wf}
C.~Degrande, N.~Greiner, W.~Kilian, O.~Mattelaer, H.~Mebane, T.~Stelzer,
  S.~Willenbrock, and C.~Zhang, {\it {Effective Field Theory: A Modern Approach
  to Anomalous Couplings}},  {\em Annals Phys.} {\bf 335} (2013) 21--32,
  [\href{http://arxiv.org/abs/1205.4231}{{\tt arXiv:1205.4231}}].

\bibitem{Kobach:2016ami}
A.~Kobach, {\it {Baryon Number, Lepton Number, and Operator Dimension in the
  Standard Model}},  {\em Phys. Lett.} {\bf B758} (2016) 455--457,
  [\href{http://arxiv.org/abs/1604.05726}{{\tt arXiv:1604.05726}}].

\bibitem{AguilarSaavedra:2018nen}
J.~A. Aguilar~Saavedra et~al., {\it {Interpreting top-quark LHC measurements in
  the standard-model effective field theory}},
  \href{http://arxiv.org/abs/1802.07237}{{\tt arXiv:1802.07237}}.

\bibitem{DAmbrosio:2002vsn}
G.~D'Ambrosio, G.~F. Giudice, G.~Isidori, and A.~Strumia, {\it {Minimal flavor
  violation: An Effective field theory approach}},  {\em Nucl. Phys.} {\bf
  B645} (2002) 155--187, [\href{http://arxiv.org/abs/hep-ph/0207036}{{\tt
  hep-ph/0207036}}].

\bibitem{Ball:2017nwa}
{\bf NNPDF} Collaboration, R.~D. Ball et~al., {\it {Parton distributions from
  high-precision collider data}},  {\em Eur. Phys. J.} {\bf C77} (2017), no.~10
  663, [\href{http://arxiv.org/abs/1706.00428}{{\tt arXiv:1706.00428}}].

\bibitem{Ball:2010gb}
{\bf The NNPDF} Collaboration, R.~D. Ball et~al., {\it {Reweighting NNPDFs: the
  W lepton asymmetry}},  {\em Nucl. Phys.} {\bf B849} (2011) 112--143,
  [\href{http://arxiv.org/abs/1012.0836}{{\tt arXiv:1012.0836}}].

\bibitem{Ball:2011gg}
R.~D. Ball, V.~Bertone, F.~Cerutti, L.~Del~Debbio, S.~Forte, et~al., {\it
  {Reweighting and Unweighting of Parton Distributions and the LHC W lepton
  asymmetry data}},  {\em Nucl.Phys.} {\bf B855} (2012) 608--638,
  [\href{http://arxiv.org/abs/1108.1758}{{\tt arXiv:1108.1758}}].

\end{thebibliography}
\end{document}